\documentclass[aps,prb,twocolumn,superscriptaddress]{revtex4}
\usepackage{graphicx}
\newcommand{\rv}{$\rho_v$}
\newcommand{\xigl}{$\xi_{\rm GL}$}
\newcommand{\xv}{$\xi_v$}
\newcommand{\msr}{$\mu$SR}

\newcommand{\lscox}{La$_{2-x}$Sr$_x$CuO$_4$}
\newcommand{\lsco}{La$_{1.85}$Sr$_{0.15}$CuO$_4$}

\begin{document}


\title{Expansion of Vortex Cores by Strong Electronic Correlation in La$_{2-x}$Sr$_x$CuO$_4$ at Low Magnetic Induction}


\affiliation{}

\author{R. Kadono}
\altaffiliation[Also at ]{ School of Mathematical and Physical Science, 
The Graduate University for Advanced Studies}
\affiliation{Institute of Materials Structure Science, High Energy Accelerator Research Organization (KEK), Tsukuba, Ibaraki 305-0801, Japan}
\author{W. Higemoto}
\affiliation{Institute of Materials Structure Science, High Energy Accelerator Research Organization (KEK), Tsukuba, Ibaraki 305-0801, Japan}
\author{A. Koda}
\affiliation{Institute of Materials Structure Science, High Energy Accelerator Research Organization (KEK), Tsukuba, Ibaraki 305-0801, Japan}
\author{M. I. Larkin}
\affiliation{Physics Department, Columbia University, New York, NY10027, USA}
\author{G. M. Luke}
\altaffiliation[Present address: ]{Department of
 Physics and Astronomy, McMaster University, Hamilton, ON L8P4N3, Canada}
\affiliation{Physics Department, Columbia University, New York, NY10027, USA}
\author{A.T. Savici}
\affiliation{Physics Department, Columbia University, New York, NY10027, USA}
\author{Y. J. Uemura}
\affiliation{Physics Department, Columbia University, New York, NY10027, USA}
\author{K. M. Kojima}
\affiliation{Department of Superconductivity, University of Tokyo, Tokyo 113-8656, Japan}
\author{T. Okamoto}
\affiliation{Department of Superconductivity, University of Tokyo, Tokyo 113-8656, Japan}
\author{T. Kakeshita}
\affiliation{Department of Superconductivity, University of Tokyo, Tokyo 113-8656, Japan}
\author{S. Uchida}
\affiliation{Department of Superconductivity, University of Tokyo, Tokyo 113-8656, Japan}
\author{T. Ito}
\affiliation{National Institute of Advanced Industrial Science and Technology, Tsukuba, Ibaraki 305-8562, Japan}
\author{K. Oka}
\affiliation{National Institute of Advanced Industrial Science and Technology, Tsukuba, Ibaraki 305-8562, Japan}
\author{M. Takigawa}
\affiliation{Department of Physics, Okayama University, Okayama 700-8530, Japan}
\author{M. Ichioka}
\affiliation{Department of Physics, Okayama University, Okayama 700-8530, Japan}
\author{K. Machida}
\affiliation{Department of Physics, Okayama University, Okayama 700-8530, Japan}


\date{\today}

\begin{abstract}
The vortex core radius \rv, defined as the peak position
of the supercurrent around the vortex, has been determined by muon
spin rotation measurements in the mixed state of \lscox\ for $x=0.13$, 0.15, 
and 0.19.
At lower doping ($x=0.13$ and 0.15),
\rv($T$) increases with decreasing 
temperature $T$, which is opposite to the behavior predicted by the 
conventional theory. Moreover, \rv($T\rightarrow0$) is significantly larger 
than the Ginsburg-Landau coherence length 
determined by the upper critical field, and shows a clear tendency to 
decrease with increasing the doping $x$.
These features can be qualitatively reproduced in a microscopic model 
involving antiferromagnetic electronic correlations.
\end{abstract}


\maketitle




The structural and electronic properties of the flux line lattice (FLL)
state in high-$T_c$ cuprates 
have been extensively studied by microscopic techniques, including 
scanning tunneling spectroscopy (STS), small angle neutron scattering (SANS),
nuclear magnetic resonance (NMR), and muon spin rotation (\msr).
Rich information has been provided not only on the superconductivity itself, 
but also on the unique FLL state as a `vortex matter' realized 
in YBa$_2$Cu$_3$O$_{7-\delta}$ (YBCO)
and Bi$_2$Sr$_2$CaCu$_2$O$_{8+\delta}$ (BSCCO) 
systems. In particular, recent developments concerning the 
\msr\ experimental technique have
made it feasible to obtain microscopic details of the spatial field
distribution $B({\bf r})$ in the FLL state directly from the 
\msr\ time spectra \cite{Sonier:00}.
The application of this advanced \msr\ technique to various 
type II superconductors including YBCO has revealed that
the vortex core radius $\rho_v$, which must be proportional to the 
Ginzburg-Landau (GL) coherence length 
$\xi_{\rm GL}$ near the transition temperature ($T_c$), strongly depends
on the external magnetic field: $\rho_v$ exhibits a sharp increase 
with decreasing field at lower fields, while it tends to converge to
$\xi_{\rm GL}\equiv\sqrt{\Phi_0/2\pi H_{c2}}$
at higher fields (with $\Phi_0$ and $H_{c2}$ being the flux quantum
and the upper critical field, respectively) 
\cite{Sonier:99,Ohishi:02}. 
This indicates that the conventional
picture of the FLL state, in which the vortices 
are regarded as being simple arrays of 
rigid cylinders (with a radius $\xi_{\rm GL}$) containing normal electrons, 
is valid only near $T_c$, i.e., the region prerequisite 
for the application of the GL theory.  
A recent calculation based on the quasiclassical Eilenberger
theory \cite{Ichioka:99} has successfully reproduced the observed 
field dependence
of $\rho_v$ in CeRu$_2$ in which an isotropic order parameter
is realized for $H/H_{c2}\le0.5$ \cite{Kadono:01}.
On the other hand, the result of a similar calculation for anisotropic 
$d$-wave superconductors is 
far from satisfactory to explain the experimental observation in YBCO,
where $\rho_v$ exhibits a much steeper change compared with the theoretical
prediction. 

Meanwhile, similar studies on the \lscox\ (LSCO) system are still 
in a preliminary stage 
because both the STS and SANS techniques have rarely been successful 
in observing vortices in this compound.  It is only very 
recently that the SANS signal from FLL has been reported in an overdoped
LSCO with the revelation of a square FLL structure at higher 
magnetic fields \cite{Gilardi:02}.
The previous $\mu$SR studies on LSCO were mostly concerned with 
remnant magnetism over the underdoped region, 
including that near $x=1/8$ \cite{Watanabe:94}, 
or with magnetic penetration depth, $\lambda$, which was evaluated from 
the total linewidth without any detailed modeling of the microscopic 
structure of vortices \cite{Aeppli:87}.  To our knowledge, there has only been 
one attempt to derive the vortex core size, which found a relatively large cutoff
parameter assuming   
the modified London model \cite{Luke:97}.

In this work, we report on high-precision
\msr\ measurements of the magnetic field distribution in the mixed
state of \lscox\ single crystals.  We found that the field
distribution at lower fields ($H<0.4$ T) is perfectly reproduced
by a modified London model with a Lorentzian cutoff. 
The actual size of $\rho_v$ defined as the peak of
the supercurrent density deduced from Maxwell's relation was 
quite large (e.g., \rv$\simeq80$ \AA\ for $x=0.15$),  
being much larger than \xigl\ ($\simeq23$ \AA\ for $x=0.15$).
More importantly, we have found that i) \rv($T$) in LSCO with $x=0.13$ 
and 0.15 tends to increase with decreasing temperature $T$, and that 
ii) \rv(0) exhibits a monotonic decrease
with increasing doping $x$. 
While these features are qualitatively unique compared with YBCO\cite{Sonier:99R} 
and have no simple explanation within the conventional theoretical model, 
they can be reproduced 
in a model involving strong antiferromagnetic correlations competing with
superconductivity.

\begin{figure}[ht]
\includegraphics[width=0.9\linewidth]{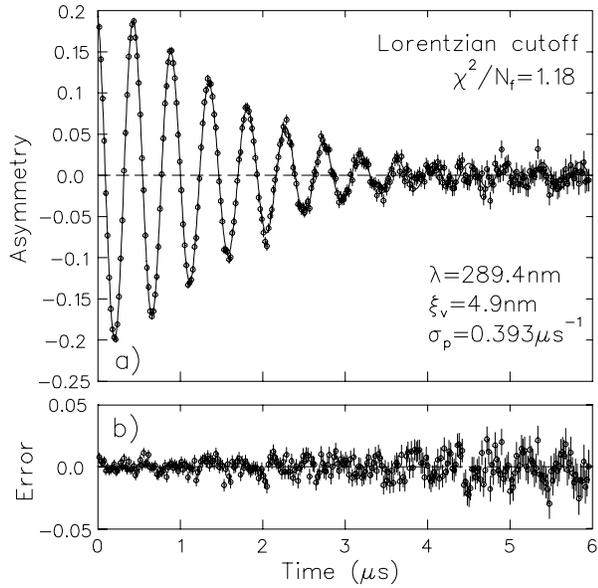}%
\caption{a) Time differential $\mu$-$e$ decay asymmetry in \lsco\
(LA15) at 15 K and $H=0.2$ T, where the solid curve shows the
result of a fitting analysis with a Lorentzian cutoff (see text). 
b) Difference between the data and the solid curve, which is defined
as the error of the fitting analysis. \label{ftspec}}
\end{figure}

\msr\ experiments on two sets of single crystalline \lscox\ (one with $x=0.15$, 
labeled LA15, and  those with $x=$0.13, 0.15, and 0.19, labeled  LB13,
LB15, and LB19, respectively)  were performed on 
the M15 beamline at the TRIUMF muon facility which provides a beam of nearly 
100\% spin-polarized positive muons of momentum 29 MeV/c. 
Some results on LB15 were previously reported \cite{Luke:97}.    
The specimen was loaded onto a He-flow cryostat 
and cooled from a temperature above $T_c$ after setting the 
magnetic field at every field point (i.e., field-cooling) 
to minimize the effect of flux pinning.  
The initial muon spin polarization was perpendicular
to the magnetic field $H$, and thus  to the FLL in the superconducting state,
as the crystal $c$-axis was alligned parallel to $H$ 
so that the field distribution
associated with the FLL was determined by the in-plane penetration depth
$\lambda_{ab}$ and the coherence length $\xi_{ab}$.
In zero field, \msr\ results from the specimen
LA15 exhibited slightly enhanced spin relaxation below $\sim$10 K,
which was virtually absent in LB15.
The spin relaxation rate in LA15 at 3.5 K
was greater than that at 41 K by 0.026(2) $\mu$s$^{-1}$, indicating
a weak remnant magnetism.  
In addition, we observed an additional
component undergoing a fast spin relaxation in LA15 in
measurements with an applied field above 0.4 T, 
which became more prominent at higher fields.
In the present study, we analyzed data obtained only for $H=0.2$ T which
do not exhibit such additional relaxation, in order to avoid any complication
due to magnetism.  
We will report on high-field \msr\ measurements 
in all of these specimens in a separate paper \cite{Savici:03}, and 
discuss the possible relation to 
field-induced magnetism suggested by neutron diffraction
\cite{Katano:00,Lake:01,Lake:02}.

\begin{figure}[ht]
\includegraphics[width=1.0\linewidth]{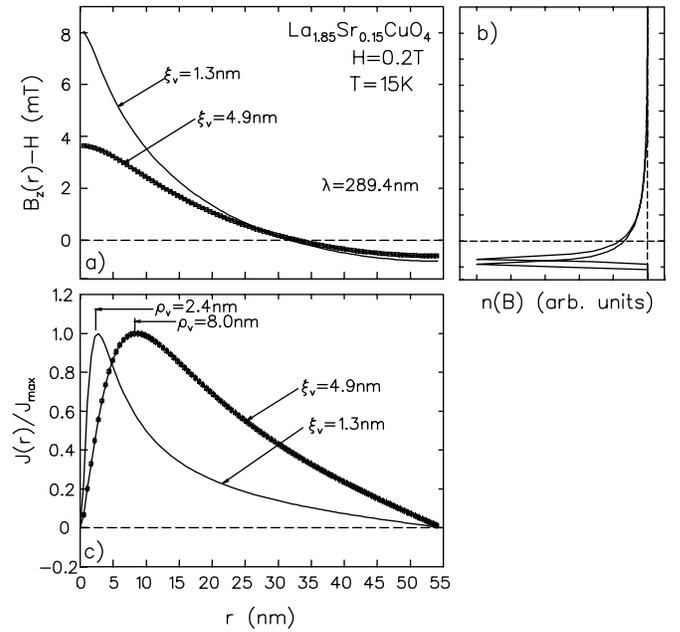}%
\caption{a) Example of the reconstructed field distribution ($B_z(r)$, 
solid line with data shown by open squares) using the result of Fig.~\ref{ftspec},
which yields the spectral density ($n(B))$ in b). The radial distribution of 
the supercurrent density, $J(r)={|\rm rot}{\bf B}({\bf r})|$, is shown in c) along 
with a definition of the core radius \rv.  For a comparison, those corresponding to
\rv=\xigl\ are also displayed. \label{fbjn}}
\end{figure}

Since muons stop randomly along the length scale of FLL, 
the time evolution of complex muon polarization $\hat{P}(t)$ provides 
a random sampling of the internal field distribution, ${\bf B}({\bf r})=(0,0,B({\bf r}))$,
\begin{eqnarray}
\hat{P}(t) &\equiv & P_x(t)+iP_y(t)\nonumber\\
&=&\exp(-\sigma_p^2t^2)\int_{-\infty}^\infty 
n(B)\exp(i\gamma_\mu Bt-i\psi)dB,\label{Pt}\\
n(B) & = & \langle\delta(B-B({\bf r}))\rangle_{\bf r},
\end{eqnarray}
where $P_{x,y}(t)$  is proportional to the time-dependent $\mu^+$-$e^+$ decay
asymmetry $A_{x,y}(t)$ deduced from the corresponding 
sets of positron counters $N_\psi(t)$ ($\psi=n\pi/2$, with $n=0,2$ and $1,3$)
after subtracting constant
backgrounds estimated from data at earlier times ($t<0$),
\begin{eqnarray}
A_x(t) &=& N_0(t)/N_\pi(t)=A_0P_x(t),\\
A_y(t) &=& \alpha N_{\pi/2}(t)/N_{3\pi/2}(t)=A_0P_y(t),
\end{eqnarray}
with $\alpha=[N_0(0)N_{3\pi/2}(0)]/[N_\pi(0)N_{\pi/2}(0)]$, $A_0$
is the normalized instrumental asymmetry, 
$\sigma_p$ is the additional relaxation due 
to random flux pinning, 
$n(B)$ is the spectral density for the internal field defined as a spatial
average ($\langle\:\rangle_{\bf r}$) of the delta function, 
$\gamma_\mu$ is the muon gyromagnetic ratio (=$2\pi\times$135.53 MHz/T),
and $\psi$  is the initial phase\cite{Brandt:88}.  
These equations indicate that the real amplitude of 
the Fourier-transformed muon
precession signal corresponds to $n(B)$ with an appropriate correction of 
$\sigma_p$.
In the modified London model, $B({\bf r})$ is approximated as the sum of
the magnetic induction from isolated vortices to yield
\begin{equation}
B({\bf r})=B_0\sum_{\bf K}\frac{e^{-i{\bf K}\cdot{\bf r}}}
{1+K^2\lambda^2}F(K,\xi_v)\:,
\label{Br}
\end{equation}
where ${\bf K}$ are the vortex reciprocal lattice vectors,
$B_0$ ($\simeq H$) is the average internal field, $\lambda$ is the
London penetration depth, and $F(K,\xi_v)$ is a nonlocal 
correction term with \xv\ being the cutoff parameter for the magnetic field
distribution; it must be stressed that $\xi_v$ is a parameter
to describe the electromagnetic response of a vortex, thereby not necessarily 
equivalent to the core radius $\rho_v$.
Considering small anisotropy predicted by the
theory \cite{Ichioka:99}, we assumed an isotropic vortex core and
associated field distribution near the vortex center. 

\begin{table}[hb]
\caption{A comparison of the central field between points above
and below the superconducting transition temperature $T_c$, where the
external field was set to 6 T above $T_c$ and then field-cooled.
The relative precision of the field ($|\Delta f/f|$) was better than
$10^{-4}$ for all cases. Note that the effect of field inhomogeneity 
and/or the Knight shift is larger at higher fields. Therefore, the above
number gives an upper bound for the uncertainty of the central field.
\label{tab1}}
\begin{ruledtabular}
\begin{tabular}{cccc}
Samples &  \multicolumn{2}{c}{$f=\gamma_\mu B_0/2\pi$ (MHz)} & $|\Delta f/f|$ \\
  &  $T>T_c$ & $T<4$ K ($\ll T_c$)  & \\
\hline
LA15 & 812.9905(22) & 812.9902(66) & $0.0(9)\times10^{-5}$ \\
LB13 & 813.0765(15) & 813.0533(41) & $2.9(5)\times10^{-5}$ \\
LB15 & 812.8405(17) & 812.7686(30) & $8.9(3)\times10^{-5}$ \\
LB19 & 813.0679(12) & 813.0384(19) & $3.6(3)\times10^{-5}$ \\
\end{tabular}
\end{ruledtabular}
\end{table}

As summarized in Table \ref{tab1}, the relative change of $B_0$ due to 
field inhomogeneity and/or the Knight shift was less than $10^{-4}$ 
throughout measurements below 50 K, and
therefore $B_0$  was fixed to the value determined in the normal state 
($T>T_c$) for the analysis of data below $T_c$. 
This well-defined $B_0$, together with the strongly asymmetric
feature of $n(B)$ against $B_0$,  allowed us to deduce the physical
parameters including $\lambda$ and $\xi_v$ without much ambiguity.
Moreover, we have found that the deduced values
of $\xi_v$ (and thereby $\rho_v$) were virtually 
independent of the apex angle ($\theta$) of FLL, while 
$\lambda$ showed a considerable dependence on $\theta$.
This feature can be readily understood by
considering the fact that the change in the apex angle has least effect on the
field distribution of single vortex near the center (corresponding to the
high-field end of $n(B)$), while it does modify the overlap of 
field distribution between the vortices. 
Based on this robustness of the analysis result against $\theta$, 
we assumed triangular FLL ($\theta=\pi/3$) throughout 
the following data analysis for simplicity.  
We have also found that the \msr\ 
spectra in LSCO at lower fields are much better reproduced by the Lorentzian
cutoff, $F(K,\xi_v)=\exp(-\sqrt{2}K\xi_v)$, compared
with the conventional Gaussian cutoff, $F(K,\xi_v)=\exp(-K^2\xi_v^2/2)$. 
Figure \ref{ftspec} shows a typical example
of the measured decay asymmetry, $A_x(t)$ ($\propto P_x(t)$), observed
at 0.2 T in specimen LA15 together with
the result of a fitting analysis.  The reduced chi-square $\chi^2/N_f$ 
(where $N_f$ denotes the number of degrees of freedom) is close to the ideal
value of $\simeq1$ when the Lorentzian cutoff is applied to Eq.~(\ref{Br}), 
while it becomes much worse as $\chi^2/N_f\simeq 1.61$ with the Gaussian cutoff.  
A similar tendency was commonly observed in all other samples.
This observation is consistent with the result of a theoretical analysis 
which showed that the Lorentzian cutoff is indeed a better approximation at 
the low-field limit \cite{Clem:75,Yaouanc:97}.  
Accordingly, all of the \msr\ time spectra were analyzed
by comparing the data with the time evolution calculated by 
Eqs.~(\ref{Pt})$\sim$(\ref{Br}) using the Lorentzian cutoff,
where $A_0$, $\psi$, $\sigma_p$, $\xi_v$, and $\lambda$ were
free parameters while $B_0$ and $\theta$ were always fixed.

While the fitting analysis of the \msr\ data is performed entirely in time domain,
one can reconstruct $B({\bf r})$ by Eq.(\ref{Br})
using the physical parameters deduced from the fitting analysis.
An example of the reconstructed field distribution, $B({\bf r})=B_z(r)$, 
along the radial direction from the core center to a saddle point is 
shown in Fig.~\ref{fbjn}a, together with the 
corresponding $n(B)$ (Fig.~\ref{fbjn}b) and supercurrent density, 
$J(r)=|{\rm rot}{\bf B}({\bf r})|$ (Fig.~\ref{fbjn}c).  
As shown in Fig.~\ref{fbjn}c, the core radius defined by
$J(\rho_v)=J_{\rm max}$ (where $J_{\rm max}$ denotes the maximum
of $J(r)$) is considerably larger than the magnetic cutoff parameter \xv, 
indicating the need for a special precaution in interpreting \xv\ directly 
as the core radius.  
However, the result of a data analysis for various fields/temperatures indicates 
that \rv\ is always proportional to \xv.
It has been estimated that $\xi_{\rm GL}\simeq 23$ \AA\ for 
an optimally doped sample ($H_{c2}\simeq 62$ T for $x=0.15$ \cite{Ando:99}).  
Provided that \rv\ is independent of the field and
determined by \xigl (i.e., \rv$\simeq0.6\sim0.8$\xigl as
predicted by theory\cite{Ichioka:99}), then \xv\ must be smaller than 
13 \AA\ to reproduce the corresponding core radius (see Fig.~\ref{fbjn}).  
However, an attempt to fit the data assuming \xv=13 \AA\ completely 
fails to reproduce the data;
the reduced chi-square $\chi^2/N_f$ was 2.76 at its best when 
other parameters (i.e., $A_0$, $\psi$, $\sigma_p$, and $\lambda$)
were set free to minimize $\chi^2/N_f$.  
This is primarily because, as evident in Fig.~\ref{fbjn}b,
the peak of $n(B)$ shifts significantly to a lower field, 
which cannot be compensated by any other parameters within the current model.
Note that $B_0$ is determined with a relative precision better than 
$\times10^{-4}$ so that the shift of the peak
in Fig.~\ref{fbjn}b ($\simeq 0.2$  mT) is readily discernible.
Thus, it is inferred from these results that \rv\ is about three
times as large as \xigl\ near the lower critical field.  This tendency is qualitatively in line 
with the theoretical prediction, although the magnitude is yet to be 
explained \cite{Ichioka:99}.

\begin{figure}[ht]
\includegraphics[width=1.0\linewidth]{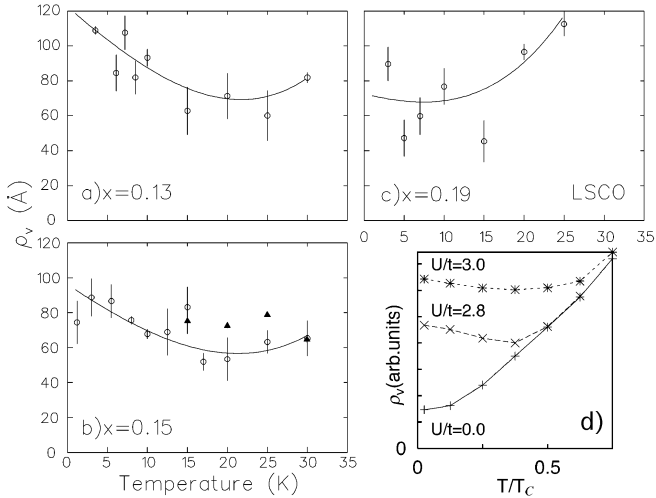}%
\caption{Temperature dependence of the vortex core radius at $H=0.2$ T
for specimens a) LB13, b) LA15 (triangles) and LB15, and c) LB19. 
d) Calculated \rv\ vs $T/T_c$ using the model in Ref.\cite{Takigawa:03}, 
where $U/t$ represents the strength of the electronic correlation 
($T_c$ is the superconducting transition temperature). The solid curves
in a)-c) show fitting results by an analitical approximation of d) to deduce the
extrapolated values $\rho_v(0)$. 
\label{frhvt}}
\end{figure}

The deduced vortex core radius for samples LB13, LB15, and LB19, 
as a function of the temperature, 
are shown in Figs.~\ref{frhvt}a-c 
together with the results
for LA15.  It exhibits a slight increase with decreasing 
temperature for $x=0.13$ and 0.15,
while an opposite tendency is observed for $x=0.19$. 
No such anomaly has been reported for YBCO \cite{Sonier:99R}.
Note that the conventional theory\cite{Kramer:74,Hayashi:98}
predicts a behavior opposite to what is actually observed 
in the former cases (see Fig.~\ref{frhvt}d). 
The values deduced from the data in LA15 are in good 
agreement with those in LB15, 
indicating that the influence of the remnant magnetism found in LA15 
is not significant at $H=0.2$ T and $T\ge15$ K.  
The penetration depth extrapolated to 0 K 
was 2559(50) \AA\ in LA15, 2446(6) \AA, 2460(4) \AA, and
2051(7) \AA\ in LB13, LB15, and LB19, respectively. The relatively large
error for LA15 is due to the influence of the remnant magnetism below 10 K. 
We also note that $\sigma_p$ was typically
0.4$\sim$0.5 $\mu$s$^{-1}$ at the lowest temperature in all specimens,
showing a common tendency of gradual decrease with increasing temperature.
This is understood as the thermal depinning of FLL from random pinning centers. 

For understanding the $T$-dependence of \rv, we performed a 
model calculation
based on Bogoliubov-de Gennes theory, where both the $d$-wave superconductivity and 
the spatially modulated antiferromagnetic (AF) spin correlations are 
simultaneously 
considered by incorporating a pairing interaction to the standard Hubbard 
model for two-dimensional square lattice\cite{Takigawa:03}.  As can be seen in Fig.~\ref{frhvt}d, the core radius 
at which the superconducting order parameter reaches 60 \% of the bulk value,
is strongly enhanced when the electronic 
correlation is present (i.e., $U/t>0$, with $U$ and $t$ being the 
respective on-site Coulomb energy and transfer matrix
element to the nearest neighboring sites in the Hubbard model),
while it obeys the prediction by Kramer and Pesch when 
$U/t=0$ \cite{Kramer:74,Hayashi:98}. 
The $T$-dependence of \rv\ for $U/t>0$ 
qualitatively agrees with those observed for $x=0.13$ and 0.15.
In the model, this feature is due to  AF spin correlations induced at the core sites, 
leading to a reduction of the local pairing amplitude and an associated increase of \rv\ at
lower temperatures.  The absence of 
quasistatic antiferromagnetism 
may be explained by the strongly dynamical spin fluctuations
in the relevant field/temperature range. 

\begin{table}
\caption{Parameter values for Eq.~(\ref{vcr}) 
deduced by fitting data in Fig.~\ref{frhvt}a--\ref{frhvt}c.
\label{tab2}}
\begin{ruledtabular}
\begin{tabular}{cccccc}
Samples &  $\rho_v(0)$ (\AA) & $c_1$ & $c_2$ & $c_3$ & $\chi^2/N_f$\\
\hline
LB13 & 120.9 (4.6) & $-$6.36(3) & $-$2.69(4) & 2.65(2) &  7.5/5 \\
LB15 & 95.0(7.3) & $-$6.42(5) & $-$2.68(6) & 2.72(4) & 9.8/7 \\
LB19 & 73.3(11.5) & $-$5.51(6) & $-$2.14(9) & 2.91(4) & 18.5/3 \\
\end{tabular}
\end{ruledtabular}
\end{table}

\begin{figure}[ht]
\includegraphics[width=1.0\linewidth]{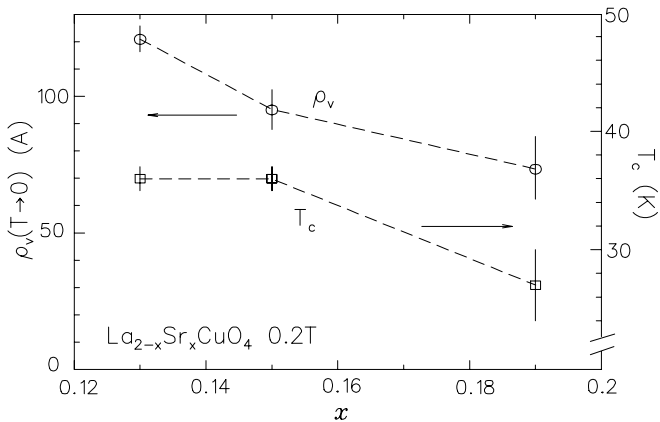}%
\caption{Doping dependence of \rv\ extrapolated 
to 0 K at $H=0.2$ T and that of the superconducting transition temperature
($T_c$) determined by the Meissner effect. The lines are guides for the eye.
\label{frhvx}}
\end{figure}

In order to obtain the extrapolated value $\rho_v(0)$, the data in Figs.~\ref{frhvt}a-3c 
were analyzed by the $\chi^2$-minimization method using an analytical 
approximation of Fig.~\ref{frhvt}d,
\begin{equation}
\rho_v(\overline{T})=\rho_v(0)\{1+c_1(\overline{T})+
c_2(\exp(-\overline{T})-1)+c_3(\exp(\overline{T})-1)\} \label{vcr}
\end{equation}
under a condition $c_1<c_2<0<c_3$ with $\overline{T}=T/T_c$.  The 
result of fitting analysis is summarized in Table \ref{tab2}.
The values of $\rho_v(0)$ are plotted against the 
Sr concentration $x$ in Fig.~\ref{frhvx}. (The data of LA15 were not
used for this analysis because they were available 
only for $T\ge15$ K, leading to large uncertainty in the fitting.) 
As the hole doping progresses with increasing $x$, the core size 
decreases monotonically.  We note that almost equivalent result was obtained
by an independent analysis using  a linear function
$\rho_v(\overline{T})=\rho_v(0)(1+c\overline{T})$ for $T<0.5T_c$.
We shall compare this behavior with those of other relevant 
parameters.  
Firstly, \rv(0) and $T_{c}$ have no simple
correlations, since $T_{c}$ shows a maximum at $x$ = 0.15.  
Secondly, the superconducting gap, $\Delta$, does not
scale with \rv(0), if $\Delta$ scales with $T_{c}$.
In contrast, \rv(0) in YBCO is larger for the lower $T_{c}$, thus following
the simple scaling law\cite{Sonier:99R}.
It has been noticed in LSCO that the gap
(or pseudo-gap) increases with decreasing hole concentration in the 
optimum-to-underdoped region.
In this sense, the (pseudo-) gap value and \rv(0) vary along with each other.
This behavior is opposite to that expected in a simple BCS
superconductor, where the coherence length is inversely
proportional to the gap value as
$\xi_0=\hbar v_F/\pi\Delta$, where $v_{F}$ denotes the Fermi velocity.
In \lscox, the Fermi velocity has been known to have little
dependence on $x$. 
It is interesting to note that the model calculation shown in Fig.~\ref{frhvt}d
appears to simulate the doping dependence if we assume that the incipient
AF correlations become weaker  ($U/t\rightarrow 0$) as the doping proceeds.
A similar result has been reported
for the correlation length, $\xi_{\rm Zn}(x)$, in Zn-doped \lscox\ 
over which the pairing is suppressed\cite{Nakano:98}.  

Finally, we point out the potential link between our results and 
those of recent neutron scattering in LSCO
where a field-induced quasistatic antiferromagnetism has been
suggested \cite{Katano:00,Lake:01,Lake:02}.
Although it has not been clearly identified as being due to the vortex cores,
they found that a long-range AF correlation is recovered 
in the mixed state under moderate magnetic fields of a few Tesla. 
A similar situation is also suggested in YBCO\cite{Mitrovic:01,Miller:02}, 
BSCCO\cite{Hoffman:02}, and YBa$_2$Cu$_4$O$_8$\cite{Kakuyanagi:02}. 
The enhanced vortex core radius
at lower fields in LSCO may be interpreted as a precursor of such a 
quasistatic correlation. 
Besides our model calculation, theories including those based on the 
$t$-$J$ model \cite{Ogata:99,Ogata:02} 
or SO(5) symmetry \cite{Zhang:97} also predict the development of AF 
correlations in place of a superconducting order
parameter when the latter is suppressed by a magnetic field or
impurity atoms.   When the superconducting order parameter
is suppressed at the center of vortices, one would expect
the emergence of AF spin correlations around the vortex cores,
which may be strongly dynamical at lower fields.

In summary, we have found in LSCO that the vortex core radius at 0.2 T is about three
times as large as that estimated from $H_{c2}$.  
The core size increases monotonically with decreasing Sr doping with an unusual 
temperature dependence for $x=0.13$ and 0.15.  Compared with the cases of
other cuprates, these features more strongly suggest a possibility 
that \rv\ is influenced by the two-dimensional AF correlations in this system.

We would like to thank the staff of TRIUMF for their technical
support during the experiment.
This work was partially supported by a
Grant-in-Aid for Scientific Research on Priority Areas and
a Grant-in-Aid for Creative Scientific Research from the Ministry of
Education, Culture, Sports, Science and Technology of Japan,
NSF Grand DMR-01-02752 and CHE-01-17752 at Columbia, and 
by NSERC of Canada.

\end{document}